\documentclass[unnumsec,webpdf,contemporary,large]{oup-authoring-template}%

\usepackage[utf8]{inputenc}
\usepackage{newunicodechar}
\newunicodechar{Δ}{$\Delta$}
\usepackage{graphicx}
\usepackage{subfigure}
\usepackage{float}
\usepackage{caption}
\usepackage{multirow}
\usepackage{makecell}
\graphicspath{{Fig/}}
\usepackage{threeparttable}
\usepackage{booktabs}
\usepackage{ulem}

\theoremstyle{thmstyleone}%

\theoremstyle{thmstyletwo}%
\theoremstyle{thmstylethree}%

\newcommand{\eg}{\textit{e}.\textit{g}.}
\newcommand{\ie}{\textit{i}.\textit{e}.}

\begin{document}

\journaltitle{Journal Title Here}
\DOI{DOI HERE}
\copyrightyear{2022}
\pubyear{2019}
\access{Advance Access Publication Date: Day Month Year}
\appnotes{Paper}

\firstpage{1}


\title[Short Article Title]{
Sequence-Only Prediction of Binding Affinity Changes: A Robust and Interpretable Model for Antibody Engineering
}

\author[1]{Chen Liu}
\author[1]{Mingchen Li}
\author[1]{Yang Tan}
\author[1]{Wenrui Gou}
\author[1,$\ast$]{Guisheng Fan}
\author[2,$\ast$]{Bingxin Zhou}

\authormark{Author Name et al.}

\address[1]{\orgdiv{School of Information Science and Engineering},
\orgname{East China University of Science and Technology}, \orgaddress{\street{Shanghai}, \postcode{200237},  
\country{China}}}

\address[2]{\orgdiv{Institute of Natural Sciences}, \orgname{Shanghai Jiao Tong University}, \orgaddress{\street{shanghai}, \postcode{200240}, \country{China}}}
\corresp[$\ast$]{Corresponding author. \href{email:email-id.com}{bingxin.zhou@sjtu.edu.cn; gsfan@ecust.edu.cn}}




\abstract{
\textbf{Motivation:}
A pivotal area of research in antibody engineering is to find effective modifications that enhance antibody-antigen binding affinity. Traditional wet-lab experiments assess mutants in a costly and time-consuming manner. Emerging deep learning solutions offer an alternative by modeling antibody structures to predict binding affinity changes. However, they heavily depend on high-quality complex structures, which are frequently unavailable in practice. Therefore, We propose ProtAttBA, a deep learning model that predicts binding affinity changes based solely on the sequence information of antibody-antigen complexes. \\
\textbf{Results:}
ProtAttBA employs a pre-training phase to learn protein sequence patterns, following a supervised training phase using labeled antibody-antigen complex data to train a cross-attention-based regressor for predicting binding affinity changes. We evaluated ProtAttBA on three open benchmarks under different conditions. Compared to both sequence- and structure-based prediction methods, our approach achieves competitive performance, demonstrating notable robustness, especially with uncertain complex structures. Notably, our method possesses interpretability from the attention mechanism. We show that the learned attention scores can identify critical residues with impacts on binding affinity. This work introduces a rapid and cost-effective computational tool for antibody engineering, with the potential to accelerate the development of novel therapeutic antibodies.
\\
\noindent \textbf{Availability and implementation:} Source codes and data are available at \url{https://github.com/code4luck/ProtAttBA}
}

\maketitle

\section{Introduction}
\label{sec1} 
Antibodies are vital components of the immune system. They induce responses through specific interactions with antigens characterized by binding affinities, which are central to antibody function and efficacy \citep{bib1,bib2}. Recent research highlights the effectiveness of antibody-based biotherapeutics, particularly in combating emerging infectious diseases \citep{zhang2021spike}.

Despite their intrinsic ability to interact with antigens, most therapeutic antibodies are not directly derived from nature but undergo laboratory screening and optimization to enhance binding affinity and achieve the desired therapeutic efficacy \citep{beck2010strategies, brustad2011optimizing}. This process typically requires extensive efforts in biological experiments on a massive number of antibody mutants. However, the time-consuming and labor-intensive nature of experimental determination for antibody mutants makes it infeasible to conduct exhaustive exploration \citep{bib8}. 

Alternatively, computational approaches provide rapid simulations and predictions of how mutations impact binding affinity. Existing assessments fall into two categories: implicit scoring and explicit scoring. Implicit scoring methods include many zero-shot protein deep models, which predict mutation effects of wild-type proteins without requiring training on labeled mutant data \citep{hsu2022esmif1,li2024mvsf}. These methods commonly use self-supervised learning to derive representations from protein sequences or structures. They enable fitness scoring for mutants without additional supervised training, where the fitness score is assumed to correlate positively with various protein properties, such as enzymatic activity, binding, and stability \citep{zhou2024protlgn,zhou2024mlife,cuturello2024enhancing}. Due to their independence from prior knowledge about specific proteins or assays, these methods are considered robust and well-suited for cold-start scenarios with limited or no experimental labels. However, their performance in scoring antibody-antigen binding affinity is suboptimal, possibly because these models do not account for antigen information.

\begin{figure*}[th]
\centering
\includegraphics[width=\linewidth]{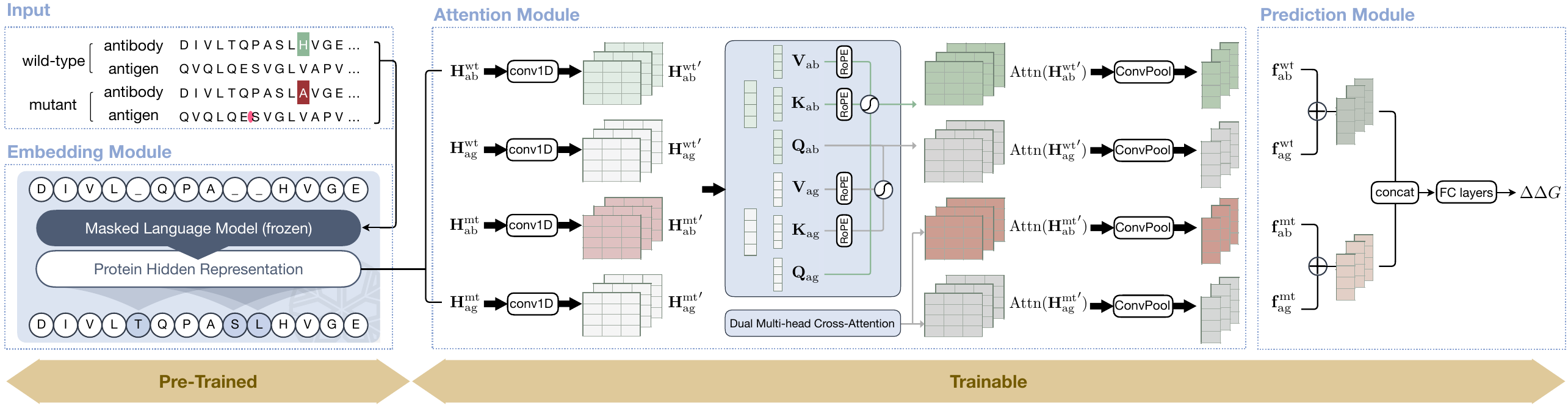}
\caption{Overview of the ProtAttBA architecture. The model predicts changes in antigen–antibody binding affinity ($\Delta\Delta G$) by amino acid mutations. Given wild-type and mutant sequence pairs, ProtAttBA first encodes antibody and antigen sequences using a frozen pre-trained protein language model to generate contextualized residue embeddings $\{\mathbf{H}_{\rm ab}^{\rm wt}, \mathbf{H}_{\rm ag}^{\rm wt}, \mathbf{H}_{\rm ab}^{\rm mt}, \mathbf{H}_{\rm ag}^{\rm mt}\}$. The attention module then applies convolutional neural networks with dual multi-head cross-attention to yield refined representations $\{\mathbf{H}_{\rm ab}^{\rm wt^\prime}, \mathbf{H}_{\rm ag}^{\rm wt^\prime}, \mathbf{H}_{\rm ab}^{\rm mt^\prime}, \mathbf{H}_{\rm ag}^{\rm mt^\prime}\}$ and the corresponding pooled feature vectors $\{\mathbf{f}_{\rm ab}^{\rm wt}, \mathbf{f}_{\rm ag}^{\rm wt}, \mathbf{f}_{\rm ab}^{\rm mt}, \mathbf{f}_{\rm ag}^{\rm mt}\}$ (see Sections~\ref{1_subec2.2} and \ref{2_subec2.2}). Finally, the prediction module concatenates wild-type and mutant features and regresses the $\Delta\Delta G$ value.}
\label{model_arc}
\end{figure*}

In contrast, explicit scoring methods calculate the binding affinity changes (\eg, $\Delta\Delta G$) of antibody mutants relative to the wild type. Two major approaches have been developed for evaluating these changes in antibody-antigen complexes, including energy function calculations and data-driven prediction methods. Energy function-based methods leverage protein structural information, integrating molecular dynamics simulations and physical computations to evaluate complex interactions and affinities \citep{bib11,dehouck2013beatmusic,bib15}. These approaches offer mechanistic insights and are grounded in molecular physics, but they face limitations in processing high-throughput data and achieving high predictive accuracy. On the other hand, machine learning-based prediction methods utilize large-scale data to learn implicit patterns in protein construction and make predictions about binding affinity changes \citep{bib14,yu2024ddaffinity}. Binding is typically considered to have a stronger correlation with protein structures \citep{tan2024simple,huang2024structure}. Consequently, various models incorporate structural information and achieve strong performance in standard evaluations on open benchmarks \citep{bib41,bib13}. However, these methods heavily rely on high-quality structural data inputs. Unlike other proteins, antibodies often lack accurate structural data, with relatively low prediction accuracy and confidence for inferred structures. When only sequence information is available (a common scenario in antibody engineering), structure-based methods demonstrate limited robustness, making their predictions less reliable. Some sequence-based methods address this issue by incorporating multiple sequence alignments (MSA) to capture amino acid co-evolutionary relationships and reduce dependence on structural data \citep{bib16}. However, reliable antibodies MSAs are often challenging to obtain, and the MSA searching during training and inference renders these models slower and unsuited for high-throughput screening \citep{tan2024rem,misra2024hiresist}.

The limitations of existing methods and the critical role of antibody engineering underscore the need for a robust and efficient tool to predict binding affinity changes in antibody-antigen complexes. Such a tool is expected to demonstrate resilience to uncertainties in input data, such as antibody structures, and efficiency in both training and inference. To address these challenges, in this study, we present ProtAttBA, a novel sequence-only method leveraging a cross-\textbf{\underline{Att}}ention mechanism for \textbf{\underline{B}}inding \textbf{\underline{A}}ffinity change prediction. As depicted in Fig.~\ref{model_arc}, ProtAttBA consists of three key components: the embedding module, the attention module, and the prediction module. (1) The embedding module processes the wild-type and mutant sequences of both antibodies and antigens, generating residue-level latent representations using pre-trained protein language models. (2) Next, the cross-attention module refines these latent representations by emphasizing information-rich features and maintaining contextual dependencies through feature transformation and integration. This step is pivotal for capturing the intricate interactions within antigen-antibody complexes, which form the foundation for the precise prediction of binding affinity changes. This module also provides interpretability for ProtAttBA by identifying and highlighting molecular interaction patterns that significantly influence binding affinities. (3) The final prediction module integrates these interaction-informed features and produces binding affinity change predictions via learnable regression heads. As demonstrated in the Results section, ProtAttBA serves as a robust and interpretable solution for predicting binding affinity changes in antibody-antigen mutants, thus fulfilling the critical demands of antibody engineering.

\section{Materials and Methods}
\label{sec2}

\subsection{2.1 Datasets}
\label{2_subsec1}
Three open benchmark datasets have been used to train and evaluate both baseline methods and ProtAttBA. AB645 \citep{bib14} and S1131 \citep{bib26} consist of single-site mutations, and AB1101 \citep{bib14} includes mutations across 1 to 7 residues. These datasets quantify changes in binding affinity using the difference in free energy ($\Delta \Delta G_{\text{bind}} = \Delta G_{\text{mut}} - \Delta G_{\text{wild}}$), where the binding free energy ($\Delta G$) was experimentally determined via surface plasmon resonance by $\Delta G = -R T \ln \left(1 / K_{\mathrm{d}}\right)$, 
with $R$ representing the universal gas constant, $T$ the absolute temperature in Kelvin, and $K_{\mathrm{d}}$ the dissociation equilibrium constant of the complex.

The experimental values for S1131 are sourced from the SKEMPI database \citep{bib25}, whereas those for AB645 and AB1101 originate from AB-bind \citep{bib24}. SKEMPI compiles 3,047 mutation-induced binding free energy changes in protein-protein heterodimeric complexes with experimentally determined structures. After redundancy removal, \cite{bib26} curated S1131 with 1,131 interface single-point mutations. Conversely, AB-bind includes 32 complexes with 7 to 246 variants per complex, all measured using consistent experimental techniques to minimize discrepancies caused by variations in experimental conditions. From this dataset, \cite{bib14} selected 645 single-point mutations as AB645 and aggregated them with 456 multi-point mutations to form AB1101. All three benchmark datasets use $\Delta \Delta G_{\text{bind}}$ (The distribution of $\Delta \Delta G_{\text{bind}}$ across the three datasets can be found in Supplementary Fig. S1.) as the prediction target, but they exhibit distinct characteristics—such as variations in mutation sites and label distributions. This difference provides a comprehensive basis for evaluating model performance across different data patterns. 

\subsection{2.2 Model Architecture}
\label{2_subsec2}
Fig.~\ref{model_arc} presents the architecture of ProtAttBA. It processes sequences of wild-type and mutant antigen-antibody complexes as input to predict the resulting change in binding affinity. The model's architecture is organized into three principal modules, operating across two conceptual phases: a pre-trained representation phase, followed by a trainable interaction and prediction phase. These modules are: an \textit{embedding module} for efficient sequence representation; an \textit{attention module} designed to capture high-dimensional interactions within the input complex; and a \textit{prediction module} that integrates features from the preceding modules to generate the final predictions. Next, we explain each of these modules in detail.

\subsubsection{2.2.1 Embedding Module}
\label{1_subec2.2}
The first embedding module takes four protein sequences as input: the wild-type antibody, the wild-type antigen, the mutated antibody, and the wild-type antigen. It generates latent representations for these sequences, for which we denote as $\{\mathbf{H}_{\rm ab}^{\rm wt}$, $\mathbf{H}_{\rm ag}^{\rm wt}$, $\mathbf{H}_{\rm ab}^{\rm mt}, \mathbf{H}_{\rm ag}^{\rm mt}\}$, which are found by a pre-trained protein language model. protein language models have demonstrated enhanced scalability and stability \citep{bib28} when applied to protein sequences. Common protein language models include BERT-style models \citep{bib43,li2024prosst}, which are better suited for predictive tasks, and GPT-style models \citep{xu2024helm,xiao2024proteingpt}, which are more appropriate for generative tasks. Here we opted for a BERT-style model, \ie a masked language model (MLM), which learns to infer the probability distribution of amino acids at masked positions based on the surrounding context, and the protein representation can serve as high-dimensional features of the protein. In empirical evaluations, we implemented four popular open-source protein language models to extract embeddings, including ProtBert \citep{bib43}, ESM1b \citep{bib21}, ESM2 \citep{bib34}, and Ankh \citep{bib22}. 

\subsubsection{2.2.2 Attention Module}\label{2_subec2.2}
The attention module processes the hidden representations of antibody-antigen complexes. Overall, the module projects the residue-level matrix representation of protein sequences to vector representations, \ie,
\begin{equation}
\label{eq:attnModule}
    \{\mathbf{H}_{\rm ab}^{\rm wt}, \mathbf{H}_{\rm ag}^{\rm wt}, \mathbf{H}_{\rm ab}^{\rm mt}, \mathbf{H}_{\rm ag}^{\rm mt}\}\rightarrow\{\mathbf{f}_{\rm ab}^{\rm wt}, \mathbf{f}_{\rm ag}^{\rm wt}, \mathbf{f}_{\rm ab}^{\rm mt}, \mathbf{f}_{\rm ag}^{\rm mt}\}.
\end{equation}
The attention module comprises three core components: a 1D convolutional operation to capture local features by modeling interactions between sequentially connected residues, a dual multi-head cross-attention to incorporate global contextual information from both the operated protein sequence and its paired sequence (\eg, if the wild-type antibody sequence is the operated protein sequence, the wild-type antigen sequence serves as the paired sequence), and a final convolutional pooling layer to compress the matrix representation into a vector representation. By integrating these components, the attention module effectively propagates both local and global interactions, ensuring a robust and comprehensive representation of antibody-antigen complex dynamics. The following sections provide detailed descriptions of each submodule. For simplicity and clarity, in this subsection we use $\mathbf{H}$ without superscripts and subscripts to denote the hidden representation of an arbitrary sequence.

\noindent \textbf{1D Convolutional Operation} \quad 
The first 1D convolutional operation learns the local patterns within the input representation space. Define 
\begin{equation}
\label{eq:1dcnn}
\hspace{-2mm} \mathbf{H}^{\prime}={\rm softmax}\left({\rm Conv1D}({\rm LayerNorm}(\mathbf{H}))\right)\odot {\rm LayerNorm} (\mathbf{H}),
\end{equation}
where ${\rm LayerNorm}(\cdot)$ denotes layer normalization to ensure numerical stability, $\odot$ represents element-wise multiplication, ${\rm softmax}(\cdot)$ is the softmax activation function, and $\mathbf{H} \in \mathbb{R}^{L\times d}$ is the $d$-dimensional embedding representation of the protein sequence with $L$ residues derived from the pre-trained language model. The ${\rm Conv1D}(\cdot)$ operator, with a kernel size of 1, calculates spatial weights for each position in the sequence. These weights are element-wise multiplied with the original representation, allowing for adaptive feature weighting and enhancing the model's sensitivity to potentially significant positions within the sequence. The same operation in \eqref{eq:1dcnn} applies to all four representations in parallel. 

\noindent \textbf{Dual Multi-head Cross-Attention} \quad
This submodule implements multi-head cross-attention to model interactions between antibody-antigen complex pairs. At this stage, following the 1D convolutional operation defined in Equation~\eqref{eq:1dcnn}, each processed representation $\mathbf{H}^{\prime}$ corresponds to an attention score matrix ${\rm Attn}(\mathbf{H^{\prime}})$. To compute the attention scores, we first define the query, key, and value matrices: $\mathbf{Q} \in \mathbb{R}^{L \times d}$, $\mathbf{K} \in \mathbb{R}^{L \times d}$, and $\mathbf{V} \in \mathbb{R}^{L \times d}$, which are parameterized by learnable weight matrices $\mathbf{W}_q$, $\mathbf{W}_k$, and $\mathbf{W}_v$, respectively:
\begin{equation} 
\label{eq:qkv}
\mathbf{Q} = {\rm RoPE}(\mathbf{W}_q \mathbf{H}^{\prime}), \; \mathbf{K} = {\rm RoPE}(\mathbf{W}_k \mathbf{H}^{\prime}), \; \mathbf{V} = \mathbf{W}_v \mathbf{H}^{\prime}. 
\end{equation}
Rotary position embedding (RoPE) \citep{bib45} is applied to enhance sensitivity to spatial relationships between residues. The same computations are applied in parallel to all four input embeddings. 

The next step computes cross-attention for antibody-antigen pairs. The calculation is performed separately for the wild-type and mutated pairs. For each pair, a symmetric operation is applied to the respective antibody and antigen components. The overall cross-attention mechanism is defined as follows:
\begin{align} 
\label{eq:crossAttn} 
{\rm Attn}(\mathbf{H}_{\rm ab}^{\prime}) &= {\rm softmax} \left( \frac{\mathbf{Q}_{\rm ab}(\mathbf{K}_{\rm ag})^{\top}}{\sqrt{d}} \right) \mathbf{V}_{\rm ag}, \\
{\rm Attn}(\mathbf{H}_{\rm ag}^{\prime}) &= {\rm softmax} \left( \frac{\mathbf{Q}_{\rm ag} (\mathbf{K}_{\rm ab})^{\top}}{\sqrt{d}} \right) \mathbf{V}_{\rm ab}. 
\end{align}
Here $d$ denotes the projection dimension associated with $\mathbf{K}$ and $\mathbf{Q}$. By integrating this symmetric cross-attention mechanism, the model facilitates effective communication between antibody and antigen sequences, thus capturing interactions in antibody-antigen complexes more comprehensively. A multi-head attention is applied capture a rich pattern representation:
\begin{equation}
\label{eq:mh}
\mathbf{H}_o=\mathbf{W}_o {\rm concat}([{\rm Attn}(\mathbf{H}^{\prime}_1),{\rm Attn}(\mathbf{H}^{\prime}_2),\dots,{\rm Attn}(\mathbf{H}^{\prime}_N)]).
\end{equation}
Here, $\mathbf{H}^{\prime}_i$ denotes the vector representation of the $i$th attention head, and $\mathbf{W}_o$ represents the learnable linear projection matrix. Same as before, this procedure is applied to all four protein representations in parallel.

\vspace{2mm}
\noindent \textbf{Convolutional Pooling} \quad 
The transformed representations $\mathbf{H}_o$ undergo a convolutional pooling and a weighted summation to derive the vector representation of each antibody or antigen
\begin{align}
\hspace{-2mm}
\mathbf{f}&=\sum_{l=1}^L({\rm softmax}(\mathbf{W}_c {\rm LayerNorm}(\mathbf{H}^{l}_o))\odot
{\rm LayerNorm}(\mathbf{H}^{l}_o)),
\label{poolfeature}
\end{align}
where $\mathbf{H}^{l}_o$ denotes the $l$th column in $\mathbf{H}_o$, \ie, the $i$th position of the protein. The same pooling operation applies to the matrix representation of all four proteins in \eqref{eq:mh}. 
After the final pooling step by \eqref{poolfeature}, we obtain four protein-level vector representations $\{\mathbf{f}_{\rm ab}^{\rm wt}, \mathbf{f}_{\rm ag}^{\rm wt}, \mathbf{f}_{\rm ab}^{\rm mt}, \mathbf{f}_{\rm ag}^{\rm mt}\}$, which will be sent to the prediction module. 

\subsubsection{2.2.3 Prediction Module}
\label{3_subec2.2}
The final prediction module integrates the joint representations of wild-type and mutant complexes to predict the binding affinity changes through fully connected layers. Based on the output of the previous step, the joint vector representation $\mathbf{f}$ is summarized by summing the information from the wild-type and mutant complexes, $\mathbf{f}={\rm concat}(\mathbf{f}_{\rm ab}^{\rm wt} + \mathbf{f}_{\rm ag}^{\rm wt}, \mathbf{f}_{\rm ab}^{\rm mt} + \mathbf{f}_{\rm ag}^{\rm mt})$.
The representation $\mathbf{f}$ is then passed through three fully connected layers to predict $\Delta\Delta G$ induced by the mutation, \ie,
\begin{align} 
\mathbf{\hat{y}} = \mathbf{W}_3 \cdot {\rm Tanh}(\mathbf{W}_2 \cdot {\rm ReLU}({\rm dropout}(\mathbf{W}_1 \cdot \mathbf{f}))),
\end{align}
where $\{\mathbf{W}_1,\mathbf{W}_2,\mathbf{W}_3\}$ are learnable parameters, ${\rm Tanh}(\cdot)$ and ${\rm ReLU}(\cdot)$ are activation functions, ${\rm dropout}(\cdot)$ denotes the dropout operations, and $\mathbf{\hat{y}}$ is the final numerical prediction, which, in our case, is $\Delta\Delta G$ of the antigen-antibody complex before and after the mutation.

\begin{table*}[ht]
\caption{Performance comparison on three open benchmarks with K-fold validation. We highlighted the \textbf{best} and \underline{second best} results.}
\label{tab1}
\centering
\resizebox{\linewidth}{!}{
    \begin{tabular}{lcccccccccccc}
    \toprule
    & \multicolumn{4}{c}{AB645} 
    & \multicolumn{4}{c}{S1131} 
    & \multicolumn{4}{c}{AB1101} \\
    \cmidrule(lr){2-5} 
    \cmidrule(lr){6-9} 
    \cmidrule(lr){10-13}
    Model  & RMSE & $R^2$ & PCC & $\rho$  & RMSE & $R^2$ & PCC & $\rho$ & RMSE & $R^2$ & PCC & $\rho$ \\
    \midrule
    \multicolumn{13}{c}{Sequence-based Methods}\\
    \midrule
    DeepEP-PPI\textsuperscript{*}  
    & - & 0.09 & 0.41  & -
    & - & 0.03 & 0.21  & -
    & - & 0.28 & 0.54  & -
    \\
    LSTM-PHV\textsuperscript{*} 
    & - & 0.07 & 0.17 & -
    & - & 0.19 & 0.39 & -
    & - & 0.05 & 0.16 & -
    \\
    PIPR\textsuperscript{*} 
    & - & 0.10 & 0.20 & -
    & - & 0.21 & 0.33 & -
    & - & 0.19 & 0.37 & -
    \\
    TransPPI\textsuperscript{*}
    & - & 0.07 & 0.18 & -
    & - & 0.19 & 0.38 & -
    & - & 0.12 & 0.22 & -
    \\
    AttABseq\textsuperscript{*}  
    & 1.75 & 0.17 & 0.44 & -
    & 1.82 &0.37 &0.66 & -
    &1.72 &0.34 &0.59 & -
    \\
    \midrule
    \multicolumn{13}{c}{Structure-based Methods}\\
    \midrule
    BeAtMuSiC 
    & 1.98$\pm$0.23 
    & -0.03$\pm$0.11 
    & 0.26$\pm$0.13
    & 0.38$\pm$0.08
    & 2.37$\pm$0.41 
    & 0.05$\pm$0.10 
    & 0.29$\pm$0.13 
    & 0.36$\pm$0.09
    & - & - & - 
    \\
    FoldX-PDB 
    & 2.51$\pm$0.78 
    & -0.83$\pm$1.11 
    & 0.31$\pm$0.20 
    & 0.29$\pm$0.12
    & 2.65$\pm$0.50 
    & -0.22$\pm$0.36
    & 0.43$\pm$0.08
    & 0.47$\pm$0.09
    & 3.40$\pm$0.48 
    & -1.66$\pm$0.78 
    & 0.28$\pm$0.11 
    & 0.26$\pm$0.14
    \\
    FoldX-AF2
     &3.04$\pm$1.41 
     &-2.02$\pm$3.33 
     &0.13$\pm$0.14
     &0.14$\pm$0.11
     &3.14$\pm$0.69 
     &-0.85$\pm$0.94 
     &0.39$\pm$0.08
     &0.49$\pm$0.08
     &3.96$\pm$0.79
     &-2.65$\pm$1.27 
     &0.14$\pm$0.07
     &0.08$\pm$0.05
     \\
    FoldX-ESM 
    &3.32$\pm$1.34 
    &-2.27$\pm$2.58 
    &0.04$\pm$0.13 
    &0.05$\pm$0.13
    &2.72$\pm$0.32 
    &-0.28$\pm$0.12 
    &0.08$\pm$0.07 
    &0.09$\pm$0.07
    &3.89$\pm$0.96 
    &-2.49$\pm$1.49 
    &0.09$\pm$0.08
    &0.04$\pm$0.06
    \\
    DDGPred-PDB  
    &\textbf{1.69$\pm$0.51} 
    &\textbf{0.25$\pm$0.23}
    &\textbf{0.54$\pm$0.16}
    & \textbf{0.62$\pm$0.11}
    &\textbf{0.95$\pm$0.13}
    &\textbf{0.84$\pm$0.04} 
    &\textbf{0.92$\pm$0.02}
    & \textbf{0.85$\pm$0.02}
    &1.79$\pm$0.16
    &0.28$\pm$0.03
    &0.59$\pm$0.02
    &0.53$\pm$0.02
    \\
    DDGPred-AF2 
    &2.19$\pm$0.29 
    &-0.34$\pm$0.33 
    &0.21$\pm$0.13
    &0.23$\pm$0.14
    &1.63$\pm$0.13
    &0.52$\pm$0.14 
    &0.76$\pm$0.07
    &0.60$\pm$0.08
    &2.37$\pm$0.11
    &-0.29$\pm$0.18
    &0.17$\pm$0.06
    &0.13$\pm$0.04
    \\
    DDGPred-ESM 
    &2.01$\pm$0.48 
    &-0.08$\pm$0.17 
    &0.37$\pm$0.19
    &0.43$\pm$0.12
    &2.39$\pm$0.30 
    &0.02$\pm$0.11
    &0.39$\pm$0.11 
    &0.37$\pm$0.08
    &2.04$\pm$0.23
    &0.06$\pm$0.05 
    &0.48$\pm$0.02
    &0.43$\pm$0.05
    \\
    \midrule
    \multicolumn{13}{c}{Ours}\\
    \midrule
    ProtAttBA-ESM2 
    & \uline{1.70$\pm$0.25}
    & \uline{0.20$\pm$0.09}
    & 0.47$\pm$0.11
    &0.48$\pm$0.12
    & 1.31$\pm$0.09
    & \uline{0.69$\pm$0.09}
    & \uline{0.84$\pm$0.05}
    & 0.75$\pm$0.06
    & \uline{1.61$\pm$0.07}
    & \uline{0.42$\pm$0.06}
    & \uline{0.65$\pm$0.04}
    & \uline{0.63$\pm$0.04}
    \\
    ProtAttBA-ESM1b
    &1.71$\pm$0.26
    &0.19$\pm$0.11
    &0.47$\pm$0.10
    &0.47$\pm$0.11
    &1.36$\pm$0.14
    &0.65$\pm$0.12
    &0.82$\pm$0.06
    &0.70$\pm$0.07
    &1.62$\pm$0.16
    &0.41$\pm$0.11
    &0.64$\pm$0.08
    &0.63$\pm$0.09
    \\
    ProtAttBA-Ankh
    &1.72$\pm$0.27
    &0.18$\pm$0.13
    &0.46$\pm$0.11
    &0.47$\pm$0.10
    &\uline{1.29$\pm$0.12}
    &0.69$\pm$0.11
    &0.84$\pm$0.06
    &0.76$\pm$0.06
    &\textbf{1.52$\pm$0.09}
    &\textbf{0.48$\pm$0.06}
    &\textbf{0.69$\pm$0.04}
    &\textbf{0.66$\pm$0.03}
    \\
    ProtAttBA-ProtBert
    &1.73$\pm$0.28
    &0.18$\pm$0.11
    &\uline{0.47$\pm$0.09}
    &\uline{0.49$\pm$0.13}
    &1.37$\pm$0.13
    &0.64$\pm$0.16
    &0.81$\pm$0.10
    &0.71$\pm$0.09
    &1.62$\pm$0.07
    &0.41$\pm$0.07
    &0.65$\pm$0.05
    &0.62$\pm$0.04
    \\
    \bottomrule\\[-2.5mm]
    \multicolumn{13}{l}{$\star$ Results are obtained from \cite{bib16}.}
    \end{tabular}
}
\end{table*}

\section{Results}
\label{sec3}
\subsection{3.1 Training and Evaluation Protocol}
\label{3_subsec1}
Our model was trained and evaluated on the three open benchmarks introduced in Section~\ref{2_subsec1}. For efficiency, we opted to freeze the pre-trained embedding module. The details of the four employed pre-trained models can be found in Supplementary Table S1. Model optimization was performed using AdamW \citep{bib35} with a learning rate of $3\times 10^{-5}$. The optimization objective was to minimize the mean squared error (MSE) between the predictions and the ground truth values. Using early stopping to avoid overfitting. All experiments were conducted on a single NVIDIA RTX-3090 GPU, and the program was based on PyTorch-2.1.2.

For evaluation, we adopted a comprehensive assessment protocol with three split strategies: (1) \textbf{K-fold cross-validation}: we followed the widely-used protocol established by \cite{bib16}, comparing average validation performance via 10-fold cross-validation for AB645 and S1131, and 5-fold cross-validation for AB1101. (2) \textbf{Sequence identity split}: sequences were clustered using MMSeqs2 at 30\% identity and then divided into training and test sets with an 8:2 ratio. This setup evaluates the model’s extrapolation to sequences with low homology. (3) \textbf{Mutation depth split}: the AB1101 dataset was split by mutation order, using single-point mutations for training and multi-point mutations for testing, to assess the model’s ability to generalize from low- to high-order mutational effects. In the subsequent analysis, we name this test case AB1101-MutDepth. For all three split settings, we assess the model performance with  Root Mean Square Error (RMSE), coefficient of determination (R$^2$), Pearson correlation coefficient (PCC), and Spearman coefficient ($\rho$).

\begin{table*}[h]
\caption{Performance comparison on three open benchmarks with sequence identity split and mutation depth split. We highlighted the \textbf{best} and \underline{second best} results.}
\label{tab2}
\centering
\resizebox{0.8\linewidth}{!}{
    \begin{tabular}{lcccccccccccc}
    \toprule
    & \multicolumn{3}{c}{AB645} 
    & \multicolumn{3}{c}{S1131} 
    & \multicolumn{3}{c}{AB1101} 
    & \multicolumn{3}{c}{AB1101-MutDepth} \\
    \cmidrule(lr){2-4} 
    \cmidrule(lr){5-7} 
    \cmidrule(lr){8-10}
    \cmidrule(lr){11-13}
    Model  & RMSE & PCC & $\rho$   & RMSE & PCC & $\rho$  & RMSE & PCC & $\rho$ & RMSE & PCC & $\rho$ \\
    \midrule
    RF Regressor  
    & 1.99 & -0.05& -0.40  
    & 1.96 & 0.61 & 0.65 
    & \uline{2.35} & 0.21 & \uline{0.47} 
    &\uline{2.39} &0.46 &0.30 
    \\
    GB Regressor
    & 5.10 & -0.32 & -0.37
    & 2.02 & 0.49 & 0.56
    & 3.41 & -0.12 & 0.09
    &5.57 &-0.01 &0.02
    \\
    AttABseq\textsuperscript{*}  
    & \textbf{1.34} & \uline{0.26} & 0.29
    & 2.33 & 0.05 & 0.11
    & 2.46 & 0.04 & -0.01
    &2.64 &0.13 &0.20
    \\
    FoldX-PDB 
    & 1.61 & \textbf{0.41} & \uline{0.35} 
    & 1.89 & 0.64 & 0.61 
    & 3.86 & \uline{0.41} & \textbf{0.48}
    &4.23 &0.34 &0.27
    \\
    FoldX-AF2
     & 2.38 & 0.16 & 0.15
     & 2.03 & 0.61 & 0.60
     & 5.06 & 0.29 & 0.31
     &4.83 &0.22 &0.09
     \\
    FoldX-ESM 
    & 1.87 & 0.11 & 0.08 
    & 2.64 & 0.02 & 0.01
    & 4.96 & 0.27 & 0.24
    & 4.42 & 0.21 & 0.10
    \\
    DDGPred-PDB\textsuperscript{*}  
    &2.26 &0.08 &0.01
    &\textbf{1.51} &\textbf{0.76} &\uline{0.76}
    &2.73 &0.13 &0.37
    & 3.01 &\uline{0.50} &0.33
    \\
    DDGPred-AF2\textsuperscript{*} 
    &3.04 &-0.09 &-0.25 
    &6.60 &0.50 &0.59
    &2.85 & -0.12 & 0.02
    &6.38 &0.27 &\uline{0.35}
    \\
    DDGPred-ESM\textsuperscript{*} 
    &2.07 &-0.04 &0.01
    &2.11 &0.51 &0.54 
    &2.82 &-0.02 &0.07
    &6.60 & 0.26 &0.26
    \\
    ProtAttBA-ESM2 
    & \uline{1.44} &\textbf{0.41}  &\textbf{0.49}
    & \uline{1.70} & \uline{0.72} & \textbf{0.77}
    &\textbf{2.11} &\textbf{0.43} &0.39
    & \textbf{2.10} &\textbf{0.55} &\textbf{0.45}

    \\
    \bottomrule\\[-2.5mm]
    \end{tabular}
}
\end{table*}

\subsection{3.2 Numerical Comparison with Baseline Models}
\label{3_subsec2}
We compared the performance of ProtAttBA with a range of sequence- and structure-based baseline models to evaluate the prediction accuracy and robustness. Sequence-based methods include: Random Forest Regressor (RF Regressor), Gradient Boosting Regressor (GB Regressor), DeepPE-PPI \citep{bib38}, LSTM-PHV \citep{bib39}, PIPR \citep{bib36}, TransPPI \citep{bib37}, and AttABseq \citep{bib16}. 
Structure-based methods include: BeAtMuSiC \citep{dehouck2013beatmusic}, FoldX \citep{bib11}, and DDGPred \citep{bib41}.

The performance comparison under the first cross-validation setup is presented in Table~\ref{tab1}, where we report the average and standard deviation across four evaluation metrics. ProtAttBA consistently outperforms all baseline methods, particularly in predicting binding affinity changes for multi-site mutations in the AB1101 dataset, which includes both single-site and multi-site mutation cases. All four ProtAttBA variants achieve significantly better results than competing models. The strong performance across four different pre-trained protein language models (ESM2, ESM-1b, ProtBert, and Ankh), as shown in the last four rows of the table, highlights the flexibility and compatibility of our framework. Among them, the ESM2-based variant achieves the highest overall performance. This may be attributed to ESM2's capacity to implicitly capture evolutionary patterns in protein sequences, offering insights that are typically derived from structural data without relying on potentially unreliable predicted structures. 

\begin{figure}[tbp]
    \centering
  \includegraphics[width=0.7\linewidth]
    {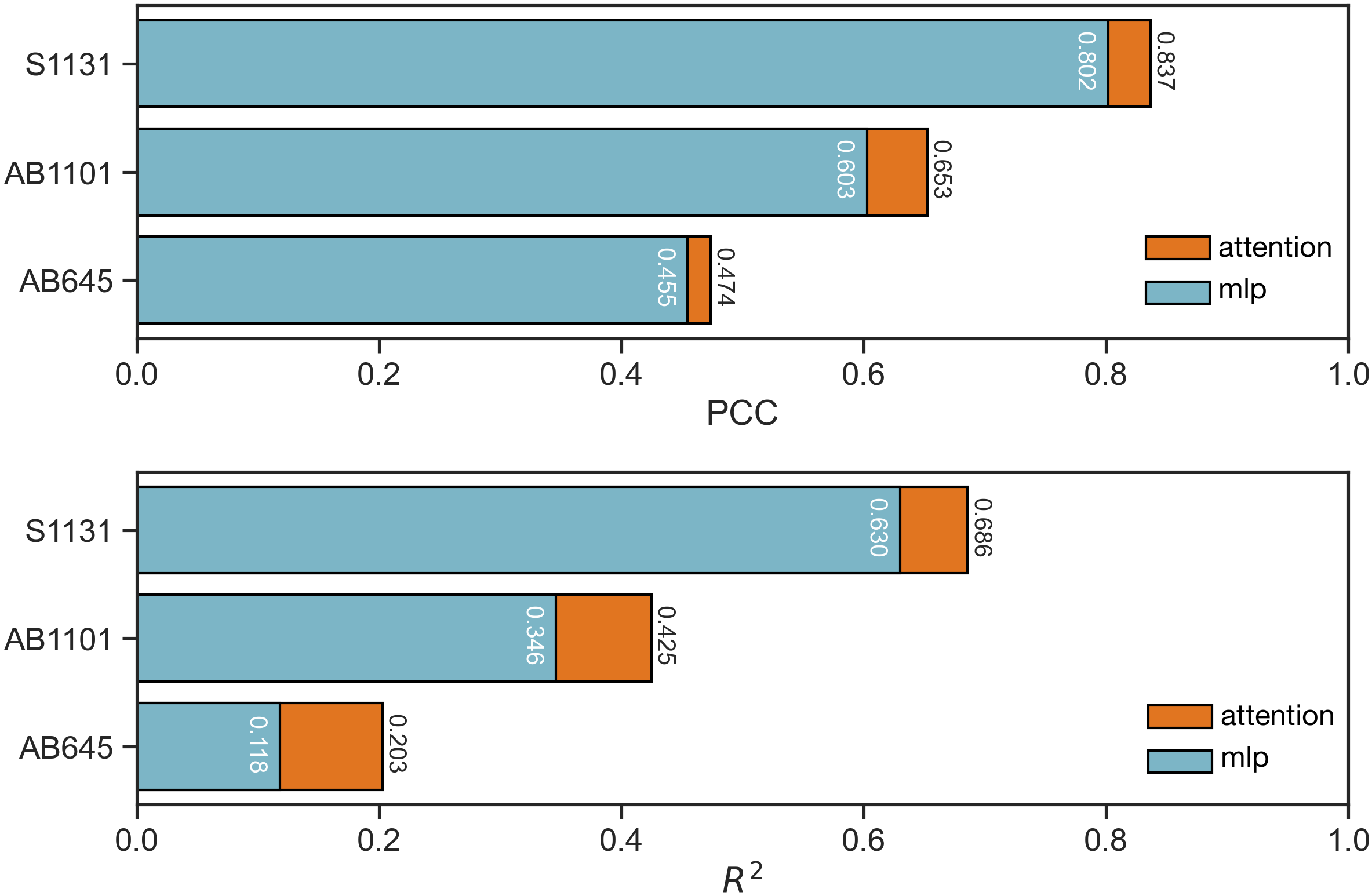}
    \caption{Ablative comparison of attention-based and MLP-based ProtAttBA by PCC (top) and $R^2$ (bottom) on the prediction performance on the three benchmark datasets.}
    \label{abla_attn}
\end{figure}

Notably, while the two structure-based deep learning methods, FoldX and DDGPred, achieve promising performance in some cases, their effectiveness is highly sensitive to input structures. PDB structures are generally considered the most accurate, followed by predicted structures from models like AlphaFold2 \citep{jumper2021highly} and ESMFold \citep{townshend2019end}. However, most proteins lack experimentally determined crystal structures. As a result, structure prediction methods like AlphaFold2 have become mainstream in protein property prediction \citep{zhou2024conditional,li2024immunogenicity}. Therefore, it is crucial for structure-based methods to be robust to input quality and still provide reliable predictions when crystal structures are unavailable. In Table~\ref{tab1}, we differentiate the performance of FoldX and DDGPred based on structure sources using the suffixes PDB/AF2/ESM, indicating crystal structures (from the dataset), AlphaFold2 (via ColabFold), and ESMFold predictions, respectively. As shown in the table, although these methods perform well on crystal structures (\eg, DDGPred-PDB ranks top on the first two datasets), their performance declines sharply with predicted structures. This observation aligns with our discussion in the introduction. That is, while structure-based models are often considered the first choice for binding-related tasks, their real-world applicability may not be as robust as sequence-based methods. 

\begin{figure*}[ht]
\centering
  \includegraphics[width=0.95\linewidth]
    {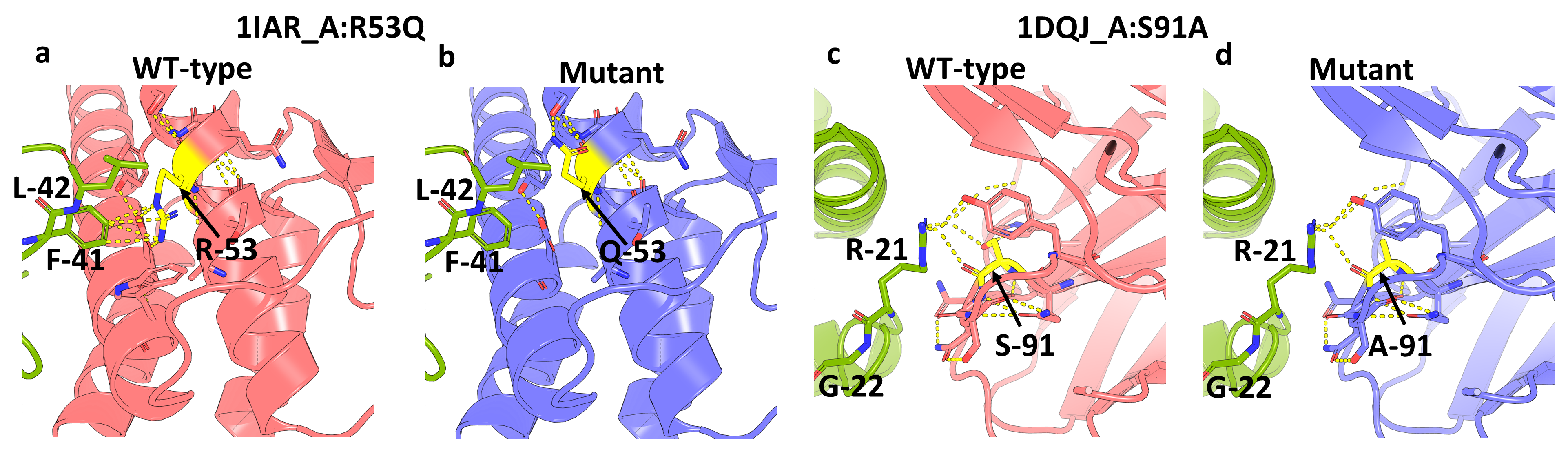} \\
    {\includegraphics[width=0.9\textwidth]{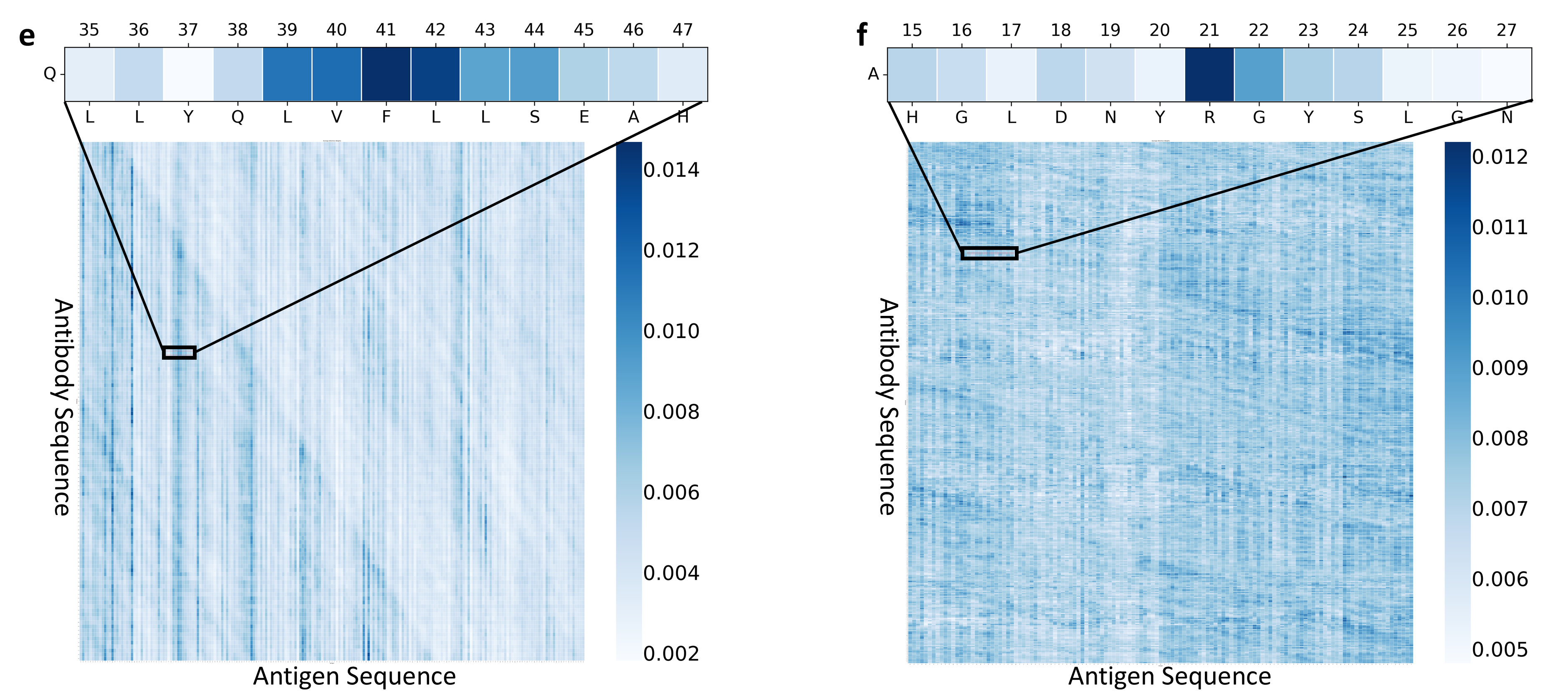}}
    \caption{Protein structure visualization for interpretability analysis. Panels a, b, c, and d depict localized views of the antibody-antigen complex at the mutation site, before and after mutation, respectively. The antigen chain is highlighted in green. Panels e and f illustrate the attention weight matrices learned by the model, where cooler colors (tending towards blue) indicate regions where the model assigns higher importance to interactions between the mutated residue and the current position.}
\label{fig_interp}
\end{figure*}

The effectiveness of ProtAttBA's attention module (intro-duced in Section~\ref{2_subec2.2}) is evaluated and reported in Fig.~\ref{abla_attn}. For a fair comparison, we replaced the multi-head attention layers with an MLP of comparable parameter count while keeping all other architectural components and hyperparameters unchanged. The consistent performance degradation observed in the MLP-based version across all datasets indicates that the attention module is crucial for enhancing the model's representation learning capabilities. This can be attributed to its ability to capture the complex interactions between antibodies and antigens, which is essential for predicting antibody-antigen binding affinity.

We further evaluated the models’ extrapolation capabilities using the sequence identity and mutation depth splits. Table~\ref{tab2} presents the performance comparison under these two evaluation protocols. We exclude the comparison of R$^2$ scores in extrapolation settings, as this metric becomes less reliable when the variance in the ground truth values increases and the prediction errors are amplified. Under the sequence identity split, most deep learning-based methods experienced a noticeable decline in performance compared to their results in the cross-validation setup shown in Table~\ref{tab1}. FoldX exhibited relatively stable and competitive performance, which can be attributed to its physics-based energy calculations that do not directly rely on learned sequence patterns from the training data. However, its strong dependence on structural accuracy remains a significant limitation--this issue becomes even more pronounced on certain benchmarks such as AB645. Meanwhile, the DDGPred variants suffered a more substantial performance drop, likely due to greater structural dissimilarity between training and testing sets after stringent sequence-based clustering. This underscores its sensitivity to both training data distribution and structural input quality. In contrast, ProtAttBA maintained relatively stable and reliable performance across different splits, suggesting that our framework generalizes more effectively to unseen proteins and is more robust to distributional shifts.

In the last four columns of Table~\ref{tab2}, we present the performance comparison under the mutation depth split on AB1101, the only dataset containing multi-site mutations. This evaluation aims to examine ProtAttBA's capability in handling complex mutational effects and its ability to generalize from simpler to more complex mutations. Overall, baseline models did not show improved performance on high-depth mutations compared to the sequence identity split. In fact, several models experienced a significant performance drop, such as FoldX, especially when relying on less accurate predicted structures as input. In contrast, ProtAttBA consistently maintained strong predictive accuracy and outperformed all baseline methods under this more challenging extrapolation setting.

In summary, across diverse evaluation scenarios we examined, ProtAttBA consistently outperformed baseline methods. The performance variability observed in structure-dependent approaches under different structure qualities and data splits highlights the robustness and broader applicability of our sequence-only framework. These results underscore ProtAttBA’s strong generalization capability and its potential as a reliable tool to predict mutation-induced changes in binding affinity, particularly in settings where structural data is unavailable or inconsistent, or when evaluating mutations in proteins with low sequence similarity.

\subsection{3.3 Model Interpretability with Attention Scores}
\label{3_subsec4}
An advantage of ProtAttBA is its ability to provide residue-level analysis of proteins, visualizing the impact of residue mutations on prediction outcomes. We randomly selected two complexes from the AB-bind and SKEMPI datasets for analysis, employing visualization techniques to examine the distribution of attention weights, focusing on mutation sites.

Fig.~\ref{fig_interp}(a-d) illustrates the hydrogen bond network alterations at mutation sites for two complexes, pre- and post-mutation. Fig.~\ref{fig_interp}a and Fig.~\ref{fig_interp}b depict the arginine (R) to glutamine (Q) mutation at position 53 of the antibody chain in 1IAR, while Fig.~\ref{fig_interp}c and Fig.~\ref{fig_interp}d show the serine (S) to alanine (A) mutation at position 91 of the antibody chain in 1DQJ. Both mutants exhibit a marked reduction in hydrogen bonds post-mutation, potentially leading to changes in antibody-antigen binding affinity. The visualization of attention weights in Fig.~\ref{fig_interp}e-f reveals that ProtAttBA identifies key amino acid positions within the hydrogen bond network. For the 1IAR position 53 mutation, strong interactions are observed with positions phenylalanine-41 (F-41) and leucine-42 (L-42) of the antigen chain. Similarly, for the 1DQJ position 91 mutation, significant interactions are noted with positions arginine-21 (R-21) and glycine-22 (G-22) of the antigen chain. These findings demonstrate the reliability of the attention-enhanced model in predicting affinity changes induced by point mutations.

\section{Conclusion}
\label{sec4}
This study addresses the critical task of predicting binding affinity changes in antibody–antigen complexes upon mutation, which plays a central role in antibody engineering. We introduce ProtAttBA, a novel deep learning framework that combines pre-trained protein language models and attention mechanisms to model protein features and interaction contexts from sequences alone. Unlike traditional structure-based methods, ProtAttBA avoids reliance on structural inputs, which are often unavailable or of uncertain quality, thus enhancing its robustness and real-world applicability.

We conducted a comprehensive evaluation of ProtAttBA under various experimental settings, including standard cross-validation, sequence identity-based splits, and mutation depth-based extrapolation. While structure-based methods demonstrate strong performance when high-quality crystal structures are available, their predictive accuracy drops significantly when fed predicted structures from folding models such as AlphaFold2 or ESMFold. This sensitivity to structural inputs limits their reliability in practical scenarios where such ideal data is rarely available (especially for antibodies and protein complexes). In contrast, ProtAttBA maintained stable performance across all settings, demonstrating strong generalization capabilities even in extrapolative scenarios, such as predicting the effects of multi-site mutations or mutations in proteins with low sequence similarity to the training data. Moreover, ProtAttBA offers potential interpretability with residue-level attention scores, allowing users to identify amino acid positions with a strong influence on binding affinity changes. These attention patterns show promising alignment with known functional sites, providing mechanistic insights and enhancing the model's transparency.

While this study emphasizes the strengths of sequence-based approaches, we do not discount the value of structural modeling. Structure-based methods remain essential, particularly for tasks where spatial configuration plays a dominant role. However, our findings highlight the need for more robust integration of structural, sequence, and evolutionary information to mitigate sensitivity to imperfect structure inputs. Future research could explore hybrid models that incorporate predicted or partial structural features, binding site annotations, or contrastive learning strategies focused on antibody–antigen interfaces to further enhance model reliability. 

\section{Competing interests}
No competing interest is declared.

\section{Author contributions statement}

Conceptualization of this study: C. L.; Implementation of the methodology: C. L.; Data curation: C. L. and G.W. R.; Investigation of the study: C. L. and B.X. Z.; Project administration: B.X. Z. and G.S. F.; Writing the paper: C. L. and B.X. Z.; Review and editing the paper: M.C. L., Y. T., and B.X. Z.; Supervision: B.X. Z. and G.S. F.
\section{Acknowledgments}
This work was supported by the grants from the National Science Foundation of China (Grant Number 62302291).

\bibliographystyle{IEEEtranN}
\bibliography{main}



\end{document}